\documentclass[12pt,preprint]{aastex}

\slugcomment{Submitted to The Astrophysical Journal}

\shorttitle{IRAS$\,$18566+0408}
\shortauthors{Araya, et al.}

\def \Lo {$\,$L$_{\odot}$}
\def \Mo {$\,$M$_{\odot}$}
\def \kms {$\,$km s$^{-1}$}
\def \mjyb {$\,$mJy$\,$beam$^{-1}$}
\def \h{$^{\rm h}$}
\def \m{$^{\rm m}$}
\def\sec{\hbox{$.\!\!^{\rm s}$}}
\def \ad {$\arcdeg$}
\def \am {$\arcmin$}
\def \as {\hbox{$.\!\!\arcsec$}}

\begin{document}

\title{An H$_2$CO 6$\,$cm Maser Pinpointing
a Possible Circumstellar Torus in IRAS$\,$18566+0408}

\author{E. Araya\altaffilmark{1,2},
P. Hofner\altaffilmark{1,2},
M. Sewi{\l}o\altaffilmark{3},
W. M. Goss\altaffilmark{2},
H. Linz\altaffilmark{4},
S. Kurtz\altaffilmark{5},\linebreak
L. Olmi\altaffilmark{6,}\altaffilmark{7},
E. Churchwell\altaffilmark{3},
L. F. Rodr\'{\i}guez\altaffilmark{5}, \&
G. Garay\altaffilmark{8}.}

\altaffiltext{1}{New Mexico Institute of Mining and Technology, 
Physics Department, 801 Leroy Place,
Socorro, NM 87801.}
\altaffiltext{2}{National Radio Astronomy Observatory, P.O. 
Box 0, Socorro, NM 87801.}
\altaffiltext{3}{Department of Astronomy, University of 
Wisconsin--Madison, 475 North Charter Street, Madison, WI 53706.}
\altaffiltext{4}{Max--Planck--Institut f\"ur Astronomie, K\"onigstuhl 17,
D--69117 Heidelberg, Germany.}
\altaffiltext{5}{Centro de Radioastronom\'{\i}a y Astrof\'{\i}sica, 
Universidad Aut\'onoma de M\'exico,
Apdo. Postal 3-72, 58089, Morelia, Michoac\'an, Mexico.}
\altaffiltext{6}{Istituto di Radioastronomia, INAF, Sezione di Firenze,
Largo Enrico Fermi 5, I-50125 Florence, Italy.}
\altaffiltext{7}{University of Puerto Rico at Rio Piedras, Physics
Department, P.O. Box 23343, San Juan, PR 00931.}
\altaffiltext{8}{Departamento de Astronom\'{\i}a, 
Universidad de Chile, Casilla 36-D, Santiago, Chile.}

\begin{abstract}

We report observations of 6$\,$cm, 3.6$\,$cm, 
1.3$\,$cm, and 7$\,$mm radio continuum,
conducted with the Very Large Array
towards IRAS$\,$18566+0408, one of the few 
sources known to harbor H$_2$CO 6$\,$cm maser 
emission.
Our observations reveal that the emission is 
dominated by an ionized jet at cm wavelengths.
Spitzer/IRAC images from GLIMPSE support 
this interpretation, given the
presence of 4.5$\,\mu$m excess emission at 
approximately the same orientation as the
cm continuum. The 7$\,$mm emission is dominated by
thermal dust from a flattened structure almost
perpendicular to the ionized jet, thus, the
7$\,$mm emission appears to trace a torus associated 
with a young massive stellar object. 
The H$_2$CO 6$\,$cm maser is coincident with the center
of the torus-like structure.
Our observations rule out 
radiative pumping via radio continuum
as the excitation mechanism for the H$_2$CO 6$\,$cm 
maser in IRAS$\,$18566+0408.
\end{abstract}


\section{Introduction}

How massive stars ($M > 10$\Mo) form is among the most 
important unsolved problems in astrophysics.
Large observational and theoretical efforts in the last ten years
are supporting the hypothesis that massive stars form
via accretion of material through disks; however other 
possible mechanisms such as spherical accretion
and coalescence cannot be discarded, mainly
in the case of stars more massive than 20\Mo~(e.g., 
Beuther et al. 2007, Cesaroni et al. 2007,
Beltr\'an et al. 2006).
Evidence for massive star formation via accretion disks
is mainly indirect, e.g., through the detection of jets
and bipolar molecular outflows (e.g., 
Arce et al. 2007; Garay et al. 2003, 2007); however more direct evidence,
such as flattened rotating structures around young massive stellar 
objects, has been observed toward a number of objects 
(for example, see the review by Cesaroni et al. 2007).
In some cases, CH$_3$OH, H$_2$O, SiO and OH masers have also 
been reported to trace circumstellar disks around young
massive stellar objects (e.g., see reviews by Cesaroni et al. 
2007 and Fish 2007).

Although the formation mechanism of massive stars is still not
well understood, the sites where massive stars are being (or have been
recently) formed are readily identifiable by the effects
that young massive stellar objects have on their environments.
Among such effects are massive molecular outflows, the formation of 
H{~\small II} regions, hot molecular cores, 
as well as the presence of maser emission from several
molecular transitions (e.g., Churchwell 2002). The 6$\,$cm K-doublet line of 
formaldehyde (H$_2$CO) is one of the molecular transitions that has 
been detected as maser emission toward massive star forming regions. 
H$_2$CO 6$\,$cm masers have several characteristics that distinguish 
them from other astrophysical masers: (1.) 
they have been found toward a small number of regions
(e.g., Araya et al. 2007b, 2004;
Mehringer et al. 1995; Pratap et al. 1994; Forster et al. 1985), 
(2.) they are weaker than other astrophysical 
masers such as H$_2$O, OH, and CH$_3$OH masers
(e.g., Araya et al. 2007b, 2007c; Hoffman et al. 2003, 2007), 
and (3.) H$_2$CO masers have been only detected 
close to very young massive stellar objects 
(such as hot molecular cores and deeply embedded infrared 
sources, e.g., Araya et al. 2006a) and appear not to be associated
with high velocity outflows or more evolved phases of massive
star formation such as compact H{~\small II} 
regions (see review article by Araya et al. 2007a).

As pointed out by Mehringer et al. (1995), the paucity of H$_2$CO 
6$\,$cm masers suggests that the pumping mechanism of the 
masers may require quite specific physical conditions for the 
inversion, conditions that may be short lived in massive star
formation environments (see Araya et al. 2007a for a discussion
of other reasons why H$_2$CO masers may be rare). 
If this were the case, then H$_2$CO 
6$\,$cm masers could become a useful probe for detection of
specific physical conditions; however at present there exists no successful
model to explain H$_2$CO 6$\,$cm masers. Although infrared and 
collisional excitation mechanisms have been discussed in 
the literature as agents capable of inverting the 
H$_2$CO 6$\,$cm levels (see Litvak 1970, Thaddeus 1972), 
only Boland \& de Jong (1981) developed 
a model specifically intended to explain one of the known 
H$_2$CO 6$\,$cm maser sources, i.e., the maser in 
NGC$\,$7538. Their model is based on radiative pumping of the 
H$_2$CO 6$\,$cm levels by radio continuum from a background
compact (emission measure $> 10^8\,$pc$\,$cm$^{-6}$) 
H{~\small II} region. 

Araya et al. (2005) explored whether
the Boland \& de Jong (1981) model is 
capable of explaining the H$_2$CO 6$\,$cm maser in 
IRAS$\,$18566+0408, but could not draw a strong conclusion 
due to insufficient observational constraints on the radio continuum 
properties of IRAS$\,$18566+0408.
Thus, to more critically test whether a pumping mechanism
by radio continuum can explain 
the H$_2$CO maser in IRAS$\,$18566+0408, and to
explore the relation between the H$_2$CO maser and the massive
star formation process in the region, 
we conducted a multi-wavelength study of the radio continuum 
in IRAS$\,$18566+0408. In $\S$1.1 we summarize the 
properties of IRAS$\,$18566+0408, our observations and 
data reduction procedure are reported in $\S2$; the 
results and discussion of the radio continuum 
observations are presented in $\S3$. 
In $\S4$ we report our conclusions.

\subsection{IRAS$\,$18566+0408: A Massive Disk Candidate with an H$_2$CO Maser}

Located at a distance of 6.7$\,$kpc, 
IRAS$\,$18566+0408 (G37.55+0.20; Mol~83, Molinari et al. 1996) 
is a massive star forming region with 
a bolometric luminosity of $\sim 6 \times 10^4\,$\Lo,
equivalent to the luminosity of an O8 ZAMS star 
(Sridharan et al. 2002; Araya et al. 2005)\footnote{Recently,
Zhang et al. (2007) estimated a total luminosity of 
$\sim 8 \times 10^4\,$\Lo~mainly based on mid to far IR
data at $\sim 1$\arcmin~resolution.}. 
Current star formation in the region is 
evident from the presence of multiple molecular outflows, 
as well as 22$\,$GHz H$_2$O and 6.7$\,$GHz CH$_3$OH maser emission
(Beuther et al. 2002a, 2002b). 
Zhang et al. (2007) found that roughly all the 
far IR luminosity comes from a single 
compact ($<5\arcsec$) source, indicating the presence of an embedded 
massive protostar. 

IRAS$\,$18566+0408 is an ideal
source for the study of massive star formation because 
only a single weak radio continuum source is found in 
interferometric observations (Carral et al. 1999, Araya et al. 2005),
and thus the study of the region is less susceptible to 
confusion.
The absence of strong radio-continuum
indicates that IRAS$\,$18566+0408 is in a phase prior to 
the development of a bright ultra-compact H{~\small II} region.
IRAS$\,$18566+0408 also harbors one of the few known H$_2$CO
6$\,$cm masers in the Galaxy (Araya et al. 2005).

Zhang (2005) listed IRAS$\,$18566+0408 as a massive 
circumstellar disk candidate.
According to Zhang (2005), the massive 
disk\footnote{In this paper we use
the term {\it``massive disk''} as synonymous
of {\it ``circumstellar torus''} 
(also referred as pseudo-disks in the literature, 
Cesaroni et al. 2007). The term ``massive disk'' 
should not be confused with ``accretion disk''.} 
has a physical size of 8000$\,$AU and mass of 60\Mo.
Zhang et al. (2007) have recently published a multi-frequency
study of IRAS$\,$18566+0408. In $\S3.3$ we discuss their
results in the context of the observations reported in this
paper.

\section{Observations and Data Reduction}

We report VLA\footnote{The Very Large
Array (VLA) is operated by the National Radio Astronomy Observatory
(NRAO), a facility of the National Science Foundation operated
under cooperative agreement by Associated Universities, Inc.} 
C (6$\,$cm), X (3.6$\,$cm), K (1.3$\,$cm),
and Q (7$\,$mm) band radio continuum observations of
IRAS$\,$18566+0408. The C and X band
observations were carried out on 2005 May 18, with the VLA
in the B configuration. The K and Q band observations were conducted
on 2005 August 5 and 2003 April 25 with the VLA in the C and D 
configurations, respectively. The default VLA continuum mode 
(4IF, 50$\,$MHz per IF) was used for the observations. 

The total integration time on-source was 
approximately 1.6, 1.0, 1.9, and 0.5 hours for 
the C, X, K, and Q band observations, respectively.
The observations were conducted in phase referencing
mode, with switching cycles (source/calibrator) of
720/120, 600/150, 120/40, and 50/50 seconds for the C,
X, K, and Q band observations, respectively. 
We used the fast-switching VLA observing mode for the
K and Q band observations, given that short 
calibration cycles are required to track
rapid phase variations due to tropospheric 
instabilities.

The data reduction was conducted following the 
VLA high-frequency recipe for calibration
and standard imaging procedures using the NRAO 
package {\tt AIPS}. In Table~1 we give the 
details of the observations.

\section{Results \& Discussion}

We detected radio continuum emission from IRAS$\,$18566+0408 at
all observed wavelengths. In Figure~1 we show the images obtained in the
different bands. In the figure we also mark the position of the H$_2$CO
6$\,$cm maser (RA = 18\h59\m09\sec975$\pm$0\sec002, 
Decl. = +04\ad12\am15\as57$\pm$0\as03, J2000, Araya et al. 2005).
In Table~2 we list
the details of the radio images and the parameters of the continuum 
emission in IRAS$\,$18566+0408.

\subsection{An H$_2$CO 6$\,$cm Maser Pinpointing
a Possible Circumstellar Torus}

In Figure~2 we show the radio spectral energy distribution (SED) 
of IRAS$\,$18566+0408 based on our observations plus the 2$\,$cm detection
by Araya et al. (2005). We found that the SED is satisfactorily 
fitted by the combination of optically thin thermal dust emission plus
optically thin Bremsstrahlung radiation from an ionized jet.
The dust emission in Figure~2 was fitted assuming 
a dust temperature of 50$\,$K, i.e., approximately
the temperature of the warm dust
component reported by Sridharan et al. (2002),
and a dust opacity $\kappa_{\nu}$ proportional to $\lambda^{-1.8}$ 
(with $\kappa_{43.3\,\mathrm{GHz}} = 0.044\,$cm$^2\,$g$^{-1}$) 
which was obtained from a fit of the opacities
for dust particles with thin ice mantles and 
n$_{H_2} = 10^6\,$cm$^{-3}$ presented by
Ossenkopf \& Henning (1994). Our assumption of
$\kappa_{\nu}$ proportional to $\lambda^{-\beta}$ where
$\beta = 1.8$, is consistent with the recent empirical 
determination of $\beta$ by Hill et al. (2006) for 
dust at 50$\,$K.

Although the cm emission can be also well fit by 
free-free radiation from a simple spherical H{\small II}
region (see \S3.2), we decided to fit  
the cm emission using an ionized  
jet model given the elongated morphology 
of the 6$\,$cm and 3.6$\,$cm continuum emission (Figure~1); 
specifically, we used the ``standard'' 
collimated model of Reynolds (1986). 
The fit was obtained 
assuming an electron density of 2$\times 10^4\,$cm$^{-3}$ 
at 1340$\,$AU (i.e., 0\as2 at 6.7$\,$kpc) from the protostar, 
an electron temperature of 10$^4\,$K, fully ionized gas, 
a width at the base of the jet of 1340$\,$AU, 
and a half-length of the ionized jet 
of $6700\,$AU. In addition, we assumed that the jet is 
in the plane of the sky. The estimated ionized mass-loss rate
is greater than $10^{-6}$\Mo$\,$yr$^{-1}$ (assuming a jet
velocity of $>$100\kms), which, for example, is comparable to 
the mass-loss rate of the ionized jet in GGD$\,$27 (Mart\'{\i} 
et al. 1995).   
The overall cm-mm SED of IRAS 18566+0408 is very similar in shape, but
much more luminous, when compared to those found in some low mass young stars 
(e.g., L1551 IRS5, Rodr\'{\i}guez et al. 1998).
Since we only detected optically-thin radio emission from the jet, 
we cannot discriminate between different 
Reynold's models, i.e., by adjusting the parameters all models can 
produce good fits. Thus, we are unable to precisely determine
physical parameters of the ionized gas such as the electron
temperature and density.

The interpretation that the elongation detected at 
6$\,$cm and 3.6$\,$cm is caused by an ionized jet is strengthened
by Spitzer/IRAC data of IRAS$\,$18566+0408 from GLIMPSE. In Figure~3 we
show a three-color image of IRAS$\,$18566+0408 in the
3.6$\,\mu$m (blue), 4.5$\,\mu$m (green), 
and 8.0$\,\mu$m (red) bands. The Spitzer data show an excess at 
4.5$\,\mu$m toward the north-west of the
IRAS$\,$18566+0408 position. Excess in the 4.5$\,\mu$m 
band is a known tracer of outflows (shocked gas) in massive star forming
regions due to the contribution of 
H$_2$ emission lines (e.g., Noriega-Crespo et al. 2004, Smith et al. 2006). 
Thus, the similar orientation of the 4.5$\,\mu$m excess and the 
cm radio continuum supports the interpretation that the 
elongation observed at 6$\,$cm and 3.6$\,$cm 
is due to an ionized jet.

In Figure~4 we show a three-color image of the 
6$\,$cm (blue), 3.6$\,$cm (green), and 7$\,$mm (red) emission
as shown in Figure~1. The ionized jet is elongated in a 
$\sim$E--W orientation, and although the signal-to-noise ratio of the 
Q-band detection is low ($\sim 7\,\sigma$),
the orientation of the 7$\,$mm source 
(after beam deconvolution) is in the NE--SW direction.
Since the 7$\,$mm emission is almost 
perpendicular to the ionized jet and is
dominated by thermal dust, it appears that
the 7$\,$mm source traces a circumstellar torus 
in IRAS$\,$18566+0408.
In Figure~4 we also show the location of the H$_2$CO 6$\,$cm maser
(Araya et al. 2005). The maser is coincident
with the center of the torus-like structure.

Using the 7$\,$mm dust flux density from the fit (Figure~2), 
assuming a 7$\,$mm dust opacity of 
$\kappa_{43.3\,\mathrm{GHz}} = 0.044\,$cm$^2\,$g$^{-1}$,
a gas-to-dust ratio of 100, and a dust temperature of 50$\,$K, 
the mass of the 7$\,$mm source is $\sim$500\Mo~(if the gas
temperature is 100$\,$K, the mass would be a factor of 2 smaller). 
Thus, the 7$\,$mm source is indeed tracing a {\it massive} 
flattened structure. This torus candidate appears to be more 
massive and greater than the parameters reported
by Zhang (2005; see however $\S3.3$). Our mass determination
should be considered a rough estimate because of the 
uncertainties in the
assumed gas-to-dust ratio, dust opacity, and dust temperature 
values. Nevertheless, we find that the mass of the 7$\,$mm source 
must be larger than 10\Mo, since even after assuming an 
extremely large 7$\,$mm dust opacity 
($\kappa_{43.3\,\mathrm{GHz}} = 0.78\,$cm$^2\,$g$^{-1}$,  
Rodr\'{\i}guez et al. 2007), 
the mass is still $>10$\Mo. 
Assuming that a single O8 ZAMS star is forming in IRAS$\,$18566+0408
and a torus mass $>10$\Mo, then the $M_{star}/M_{torus}$ ratio
for the system would be $<2$, suggesting that the
torus-like structure is dynamically unstable
(e.g., see Rodr\'{\i}guez et al. 2007, and references therein). 
Further high angular
resolution observations of the region are needed to
investigate the stability and kinematics of the 7$\,$mm source.

\subsection{Applicability of the Boland \& de Jong (1981) Model}

To test whether population inversion by radio continuum radiation
can explain the H$_2$CO maser in IRAS$\,$18566+0408, 
we adapted the Boland \& de Jong (1981) formalism
to the specific characteristics of IRAS$\,$18566+0408.
In particular, we included the contribution of dust emission to 
the photon occupation 
number\footnote{In the case of NGC$\,$7538,
Boland \& de Jong (1981) neglected dust emission
because they 
considered that free-free radio continuum radiation
was dominant at all wavelengths of interest.} and
we used the following expression for the optical depth 
of the radio continuum emission (see Reynolds 1986):

\begin{equation}
\tau_{\nu}(y) = 2 a_k w_o n_o^2 x_o^2 T_o^{-1.35} (y/y_o)^{q_\tau} \nu^{-2.1},
\end{equation}

\noindent where $a_k = 0.212$, $q_\tau = -2$ for the ``standard'' collimated
model of Reynolds (1986), $w_o = 670\,$AU, 
$n_o = 2\times 10^4\,$cm$^{-3}$, $x_o = 1$, $T_o = 10^4\,$K (see \S3.1). 
We assumed that the ionized jet is in the plane of the
sky (i.e., $sin(i) = 1$ in the Reynolds 1986 notation,
see also Figure~1 in Reynolds 1986), and hence 
$y$ is simply the distance along the ionized jet, and 
$y_o$ is the distance at which the flow is injected
(we assumed $y_o = 1340\,$AU based on the IRAS$\,$18566+0408
cm emission images, Figure~1).

Given the observational constraints imposed by our data, we find
that the model does not predict population inversion of the 6$\,$cm 
K-doublet level.
The specific ionized gas model used to fit the cm radio continuum is
not relevant to the outcome of the maser model
as long as the free-free emission
is optically thin at $\lambda \le 6\,$cm as observed.  
For example, the cm SED of IRAS$\,$18566+0408 can be fitted
by optically thin emission from a cylindrical 
H{~\small II} region characterized by n$_e = 1.7\times10^3\,$cm$^{-3}$,
T$_e = 10^4\,$K and EM$ = 1.4\times10^5\,$pc$\,$cm$^{-6}$.
No inversion is predicted for such a background
H{~\small II} region (see discussion of Figure~1 in 
Boland \& de Jong 1981, and Figure~4 in Araya et al. 2005).
We conclude that excitation by radio continuum is not a 
feasible mechanism to explain the maser in IRAS$\,$18566+0408.

\subsection{Comparison with the Observations by Zhang et al. (2007)}

Recently, Zhang et al. (2007) reported a multi-wavelength
study of IRAS$\,$18566+0408. 
They conducted 22, 43, and 87$\,$GHz
radio continuum observations, as well as 
NH$_3$ (1,1), (2,2), and (3,3), SiO (2--1), and HCN (1--0)
observations with the VLA and OVRO. In this section 
we discuss the consistency of the results found in our work 
with respect to the results found by Zhang et al. (2007).

\subsubsection{A Jet-like Outflow in IRAS$\,$18566+0408}

The SiO (2--1) observations by Zhang et al. (2007) 
reveal a well collimated molecular outflow in 
IRAS$\,$18566+0408. The outflow appears to be emanating 
from the torus-like structure reported in this work. 
The SiO outflow is in the SE--NW direction which is consistent with the
orientation of the ionized jet and the large structure outflow
traced by Spitzer (see Figures~3 and 4). However, 
the SiO outflow is not strictly collinear to the ionized jet or
the 4.5$\,\mu m$ excess emission: within $\sim 1\arcsec$ from the center
of the torus-like structure (Figure~4), the ionized jet appears
to have a position angle of $\sim 100\arcdeg$; at 
$\sim 10\arcsec$ from the torus, the position angle of the 
SiO outflow is 135$\arcdeg$; and at $\sim 1\arcmin$ scales,
the 4.5$\,\mu m$ outflow has a position angle of
$\sim 110\arcdeg$. Given that the velocity gradient of the SiO
outflow is opposite of that of the CO outflow,
Zhang et al. (2007) mentioned that the outflow may be very close
to the plane of the sky and that the velocity inversion may be
due to precession. The change in position angle between
the radio jet, the SiO and the 4.5$\,\mu m$ outflow may also be
due to precession of the jet.

The NH$_3$ and HCN data reported by Zhang et al. (2007)
show an elongation parallel to the SiO outflow, indicating
that these molecules may also be tracing the outflow. The
NH$_3$ (1,1) and (2,2) velocity integrated 
emission is distributed almost
symmetrically with respect to the center of the torus
candidate, whereas the SiO, NH$_3$ (3,3) and HCN integrated 
emission is elongated mostly toward the NW of the 7$\,$mm source.
The asymmetry of the SiO, NH$_3$ (3,3) and HCN emission
with respect to the position of the torus-like structure
is similar to the one shown by the Spitzer 4.5$\,\mu m$ 
excess, in which 4.5$\,\mu m$ excess emission is only detected 
toward the NW of the radio continuum source. 
Density gradients in the medium may be responsible for 
such an asymmetry.

\subsubsection{Continuum Observations}

Zhang et al. (2007) conducted continuum observations
at 22, 43, and 87$\,$GHz. At 87$\,$GHz ($\theta_{syn} \sim 5\arcsec$)
they detected a continuum source coincident with the
weak radio continuum source reported by Carral et al. (1999)
and Araya et al. (2005, see also Figure~1). The peak 
intensity of the emission is 18\mjyb~and the 
total flux density of the source is 31$\,$mJy. The 87$\,$GHz
flux is clearly dominated by dust emission, which is 
consistent with our radio SED interpretation. 
Zhang et al. (2007) detected 
a second mm source (MM-2) north-west from the position of the 
protostellar candidate (see their figure 1). We did not detect a
counterpart of MM-2 in our observations, which is 
consistent with the non-detection at 43$\,$GHz by
Zhang et al. (2007).

The 43$\,$GHz observations reported by Zhang et al. (2007)
were conducted with the VLA in the DnC array, resulting
on a synthesized beam of $2\as7 \times 1\as3$, which is
more elongated than our $\theta_{syn}$ at 43$\,$GHz
(Table~2). The observations of Zhang et al. (2007) have
a smaller rms than that of our observations, i.e., 
0.1 versus 0.17\mjyb. As in our observations, they also
detected 43$\,$GHz emission coincident with the 
radio continuum source. The source is slightly resolved
with a peak intensity of 1\mjyb~and flux density of
1.7$\,$mJy. The deconvolved size of the emission is 
$2\as0 \times 1\as2$, P.A. = 24$\arcdeg$, which 
corresponds to a major axis of $1.3 \times 10^4\,$AU.
Thus, the 7$\,$mm source detected by Zhang et al. (2007)
appears to be more compact and weaker than the 7$\,$mm
emission reported in this work (Table~2). To investigate
the origin of this discrepancy we reduced the 7$\,$mm
observations by Zhang et al. (2007) from the VLA archive, 
and found that the discrepancy is mainly 
due to different weighting schemes used in the imaging, i.e.,
our 7$\,$mm image was created with natural weighting to
improve signal-to-noise, whereas Zhang et al. (2007)
apparently used a more uniform weighting scheme given that
they have better sensitivity data. For example, 
if our data are imaged using a Briggs' robust 
parameter of 0, then the peak intensity of the 7$\,$mm
source is 1.0\mjyb~(i.e., the same value reported 
by Zhang et al. 2007)
and the integrated flux density is 2.6$\pm$0.6$\,$mJy.
Our reduction of the Zhang et al. data using a robust 0
weighting resulted in an integrated flux density of
1.74$\pm$0.35$\,$mJy, i.e., we reproduced the value
reported by Zhang et al. (2007) and the flux density is
consistent within the errors with the corresponding
flux density derived from our data. Regarding the size
and position angle, given the low signal-to-noise of
both data sets, and that the 7$\,$mm source is 
barely resolved, then it is difficult to accurately
measure the deconvolved size and position angle of the 
source. However, for example, if a robust 0 weighting
is used in the imaging of the Zhang et al. (2007) data,
we found that the maximum size of the 7$\,$mm source
is $3\as7 \times 1\as7$ with a position angle of 
44$\arcdeg$, which is consistent with our results
(Table~2). 
Higher sensitivity and angular resolution observations are needed to
precisely measure the size and position angle of the 7~mm source.
Nevertheless, our observations and those of Zhang et al. (2007) are
consistent with an elongated source almost perpendicular to the ionized
jet that we report in this work.

Zhang et al. (2007) also conducted VLA continuum
observations at 22$\,$GHz and found no
radio continuum emission at a $1\sigma$ level
of 0.14\mjyb. They also pointed out that 
Miralles et al. (1994) did not detect 2$\,$cm 
emission at a $1\sigma$ level of 0.16$\,$mJy whereas
Araya et al. (2005) report a 0.7$\,$mJy source
at 2$\,$cm. Zhang et al. (2007)  mentioned that variability
of the radio continuum emission in the source may
be responsible for this apparent discrepancy.
We note that Miralles et al. (1994) did not consider
a $1\sigma$ value as an appropriate detection limit
for the radio continuum, but they report upper limits
of $\le 0.5$ and $\le 0.8\,$mJy for 6$\,$cm and 
2$\,$cm emission which are consistent with our 6$\,$cm
detection (Table~2) and the 2$\,$cm detection by 
Araya et al. (2005).  
Regarding the non-detection of 22$\,$GHz 
emission reported by Zhang et al. (2007), the 5$\sigma$
detection limit of their data is 0.7\mjyb, and
the peak intensity of the 22$\,$GHz source reported in 
this work (see Figure~1) is 0.65\mjyb, i.e., at a 5$\sigma$
detection limit, the two data sets are consistent.
The non-linear nature of interferometric observations,
particularly at K band were tropospheric instabilities
may lead to phase decorrelation, can be 
responsible for the non-detection of the source at 
a 5$\sigma$ level.
Thus, given the available radio continuum data,
we find no strong evidence for variability of
the source. Nevertheless, the occurrence of flares in 
the H$_2$CO maser (Araya et al. 2007c, see also
Araya et al. 2007a) may be caused by variability
of the radio continuum, thus future observations
are required to investigate whether, and up to what
level, the radio continuum emission is variable 
in this source.

\subsubsection{Mass Determination}

Based on mm and submm wavelength data, Zhang et al. (2007)
report a spectral index $\alpha$ of 3.9 (S$_\nu \varpropto \nu^\alpha$)
and thus an emissivity index $\beta = 1.9$, 
which is consistent with the value of $\beta = 1.8$ 
derived from theoretical as well
as empirical considerations (see $\S3.1$). However,
in an effort to better constrain the value of $\beta$
by matching beam sizes, Zhang et al. (2007) derived a value of 
$\beta = 1.3$ which they believe to be more
appropriate for the dust mass determination. Based
on NH$_3$ observations, they obtained a kinetic temperature
of 80$\,$K for the molecular core associated with the 7$\,$mm, 
which was assumed to be the dust temperature. 
In addition they assumed an opacity
$\kappa(250\mu m) = 12\,$cm$^2\,$g$^{-1}$. Using 
these parameters, Zhang et al. (2007) estimated a mass of 
70\Mo~within 30000$\,$AU from the 7$\,$mm source.
Using the same parameters (T$_d = 80\,$K, 
$\kappa(250\mu m) = 12\,$cm$^2\,$g$^{-1}$, $\beta = 1.3$)
we derive a mass of $\sim 80$\Mo~based on our observations. 
Given the flux density uncertainties (Table 2) our derived mass
is consistent with the 70~M$_\odot$ mass reported by 
Zhang et al. (2007).

\subsubsection{A Molecular Core coincident with the Torus Candidate}

Zhang et al. (2007) detected compact molecular
emission from the position of the torus candidate
in their NH$_3$ and HCN observations. As mentioned above,
the NH$_3$ emission at the position of the 7$\,$mm source
traces warm ($\sim 80\,$K) material, and in addition, the line
width of the NH$_3$ emission is substantially 
broader (FWHM = 8.7\kms) in comparison
with the more extended NH$_3$ emission (FWHM $< 2$\kms). 
Zhang et al. (2007)
find it unlikely that the broad linewidth is due to the outflow,
instead the large linewidth may be due to relative motion
of multiple cores, or due to infall/rotation. They
found that if rotation is assumed, then the dynamical mass 
(assuming gravitationally bound motion) would be similar to
the 70\Mo~mass reported by Zhang et al. (2007, see $\S3.3.3$).
However, their data do not show direct kinematic evidence
for rotation of the core. 

The peak velocity of the NH$_3$ emission is 85.2\kms, which
closely agrees with the systemic velocity of the source as
traced by CS emission and H$_2$CO absorption (Bronfman et al. 1996;
Araya et al. 2004). The LSR velocity of the H$_2$CO maser 
in the region is $\sim 80$\kms, thus the maser is not tracing
gas at the systemic velocity of the molecular core but is
tracing blueshifted material, perhaps associated with the base
of the molecular outflow or a flaring disk with radial velocity
gradients.

\section{Summary} 

We report 6$\,$cm, 3.6$\,$cm, 1.3$\,$cm, and 7$\,$mm 
observations of IRAS$\,$18566+0408 
conducted with the VLA. We detected radio continuum 
emission from the source at all wavelengths. 
The cm emission appears to trace an ionized jet, whereas
the 7$\,$mm band is dominated by thermal dust emission.
The interpretation that the cm radiation arises from
an ionized jet is strengthened by Spitzer/IRAC 
observations of IRAS$\,$18566+0408, showing
excess at 4.5$\,\mu$m at the same 
orientation as the suggested ionized jet. 

The 7$\,$mm structure is elongated almost perpendicular
to the ionized jet. Taking into account that the 
7$\,$mm emission is dominated by thermal dust, then
the 7$\,$mm source appears to trace a circumstellar torus. 
The mass of the torus-like structure is greater than 
10\Mo~(and as large as several hundreds solar masses depending 
on the assumed dust opacity and temperature).
Given that definitive identification of a disk/torus requires 
kinematical evidence of rotation, future sub-arcsecond angular 
resolution molecular line observations are required 
to confirm the torus interpretation presented in this paper.

Our results are consistent with recent molecular and 
continuum observations reported by Zhang et al. (2007).
They found a SiO outflow that is approximately parallel to the
ionized jet reported in this work, and a clump
of NH$_3$ and HCN molecular gas at the position of the 
torus-like structure. Zhang et al. (2007) also conducted 7$\,$mm VLA
observations of the region and found an elongated structure
whose orientation is almost perpendicular to the 
ionized jet reported in this work, which confirms
our 7$\,$mm results.

The H$_2$CO 6$\,$cm maser in IRAS$\,$18566+0408 
(Araya et al. 2005) is coincident with the 
center of a torus-like structure which harbors a massive
protostellar candidate. This result
strengthens the association of H$_2$CO 6$\,$cm masers
with very young massive stellar objects.
The observed radio spectral energy distribution rules out pumping 
of the H$_2$CO 6$\,$cm maser in IRAS$\,$18566+0408 by radio
continuum, which Boland \& de Jong (1981) proposed as
the excitation mechanism for the H$_2$CO maser in 
NGC$\,$7538.

\acknowledgments

E.A. is supported by a NRAO predoctoral fellowship.
P.H. acknowledges support from
NSF grant AST-0454665. H.L. was supported by a postdoctoral
stipend from the German Max Planck Society.
E.C. and M.S. were supported in part by NSF grant AST-0303689.
L.O. was supported in part by the Puerto Rico Space Grant Consortium.
G.G. acknowledges support from FONDAP project No.15010003.
We thank an anonymous referee for comments that improved
the manuscript.
This research has made use of NASA's Astrophysics Data System and 
is based in part on observations made with the Spitzer Space 
Telescope, operated by the Jet Propulsion Laboratory, 
California Institute of Technology, under contract with NASA.

\clearpage

\begin{deluxetable}{lcccc}
\tabletypesize{\scriptsize}
\tablecaption{VLA Observations \label{tbl-1}}
\tablewidth{0pt}
\tablehead{
\colhead{Parameter}& \colhead{C-Band (6$\,$cm)}  & \colhead{X-Band (3.6$\,$cm)}& \colhead{K-Band (1.3$\,$cm)}& \colhead{Q-Band (7$\,$mm)} }
\startdata
\rm Date           & 2005 May 18                 & 2005 May 18                 & 2005 August 5               & 2003 April 25              \\
VLA Configuration  & B-Array                     & B-Array                     & C-Array                     & D-Array                    \\
$\nu_o\,$(GHz)     & 4.86                        & 8.46                        & 22.46                       & 43.34                      \\   
RA$^a$	           & 18\h59\m09\sec9             & 18\h59\m09\sec9             & 18\h59\m10\sec0             & 18\h59\m09\sec9            \\  
Dec$^a$            & 04\arcdeg12\arcmin10\arcsec & 04\arcdeg12\arcmin10\arcsec & 04\arcdeg12\arcmin14\arcsec & 04\arcdeg12\arcmin14\arcsec\\   
Flux Density Calib.& 3C286                       & 3C286                       & 3C286                       & 3C286                      \\
~~~~~Assumed S$_\nu$ (Jy) & 7.49                 & 5.21                        & 2.52                        & 1.45                       \\
Phase Calib.       & J1824+107                   & J1824+107                   & J1851+005                   & J1851+005                  \\
~~~~~Measured S$_\nu$ (Jy) & 0.77                & 0.72                        & 0.97                        & 0.66                       \\
\enddata
\tablenotetext{a}{Phase tracking center (J2000).}
\end{deluxetable}

\clearpage

\begin{deluxetable}{lcccc}
\tabletypesize{\scriptsize}
\tablecaption{Parameters of the Radio Continuum Images \label{tbl-2}}
\tablewidth{0pt}
\tablehead{
\colhead{Parameter}       & \colhead{C-Band (6$\,$cm)}     &  \colhead{X-Band (3.6$\,$cm)}  & \colhead{K-Band (1.3$\,$cm)}   & \colhead{Q-Band (7$\,$mm)} }
\startdata
Syn. Beam                 & 1\as3 $\times$ 1\as2           & 0\as93 $\times$ 0\as79         & 1\as4 $\times$ 1\as0           & 2\as0 $\times$ 1\as6           \\
Syn. Beam P.A.            &--27\arcdeg                     & 24\arcdeg                      & 5\arcdeg                       & 22\arcdeg                      \\
rms ($\mu$Jy$\,$b$^{-1}$) &  24                            & 21                             & 93                             & 168                            \\
S$_\nu$ (mJy)$^a$         & 0.52$\pm$0.09                  & 0.55$\pm$0.10                  & 0.80$\pm$0.18                  & 3.1$\pm$0.6                    \\
Peak RA$^a$               & 18\h59\m09\sec97$\pm$0\sec01   & 18\h59\m09\sec99$\pm$0\sec01   & 18\h59\m09\sec98$\pm$0.01      & 18\h59\m09\sec99$\pm$0\sec01   \\
Peak Decl.$^a$            & 04\ad12\am15\as49$\pm$0\as08   & 04\ad12\am15\as67$\pm$0\as08   & 04\ad12\am15\as74$\pm$0.06     & 04\ad12\am15\as7$\pm$0\as2     \\
Deconvolved Size$^a$      & 2\as9 $\times$ 0\as73          & 2\as3 $\times$ 0\as85          & $<$ 1\as5 $\times$ 1\as5       & 3\as3 $\times$ 1\as6           \\       
Deconvolved Size P.A.$^a$ & 85\ad                          & 103\ad                         & \nodata                        & 45\ad                          \\
Physical Size ($10^3\,$AU)& 19 $\times$ 5                  &  16 $\times$ 6                 &  $<$ 10 $\times$ 10            & 22 $\times$ 11                 \\
\enddata
\tablenotetext{a}{ Parameters obtained from a 2D Gaussian fit of the brightness distribution using the task {\tt JMFIT} in AIPS.}
\end{deluxetable}

\clearpage

\begin{figure}
\hspace*{-1cm}\vspace*{2cm}\includegraphics[scale=0.60, angle=90]{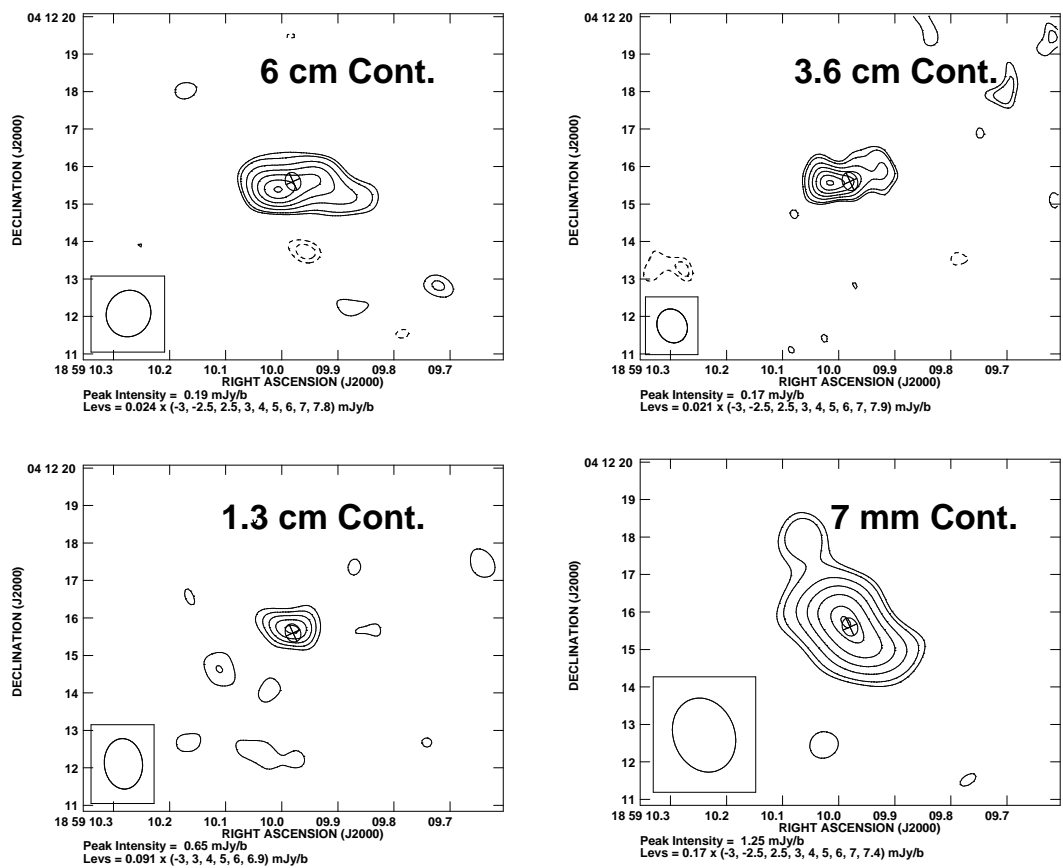}
\vspace*{-1.1cm}\caption{Radio continuum emission detected toward
IRAS$\,$18566+0408 at the four different wavelengths observed with the 
VLA. In the lower left corner of each image we show 
the respective synthesized beam (the size and position angle of the
synthesized beams are given in Table~2). 
The location of the H$_2$CO 6$\,$cm maser imaged with the 
VLA-A by Araya et al. (2005) is indicated by ``$\Earth$''.
The size and position angle of the ``$\Earth$'' symbol represents
the size and position angle of the synthesized beam of the Araya et al. (2005)
observations.}
\label{f1}
\end{figure}

\clearpage

\begin{figure}
\hspace*{0cm}\vspace*{2cm}\includegraphics[scale=0.60, angle=-90]{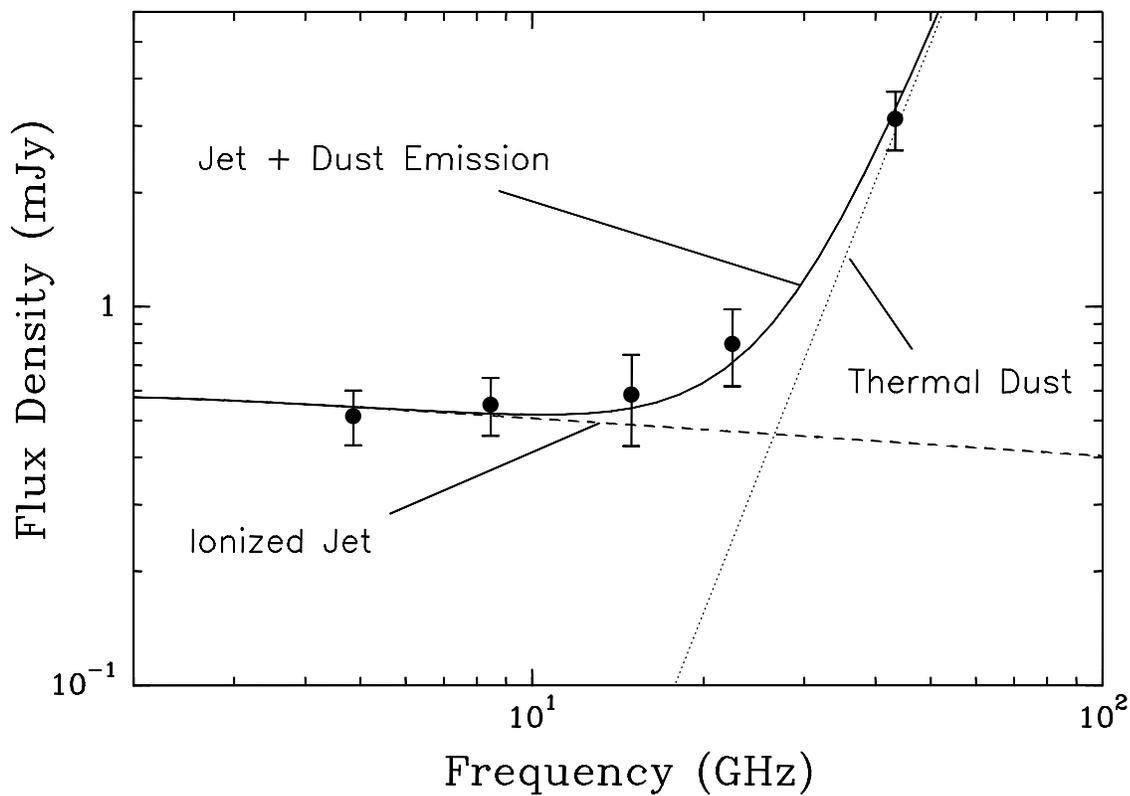}
\vspace*{-1.1cm}\caption{Radio SED of IRAS$\,$18566+0408 based on 
the observations reported in this paper (Table~2), in addition to the 
2$\,$cm data point by Araya et al. (2005). Dashed line is 
fit to the ionized gas emission using the ``standard'' collimated ionized
jet model developed by Reynolds (1986). At millimeter wavelengths the
SED is dominated by optically thin thermal dust
emission. The total fit to the SED, i.e., 
the addition of the ionized gas and thermal dust flux density, is
shown with a continuous line.}
\label{f2}
\end{figure}

\clearpage

\begin{figure}
\hspace*{-2.5cm}\vspace*{2cm}\includegraphics[scale=0.80, angle=-90]{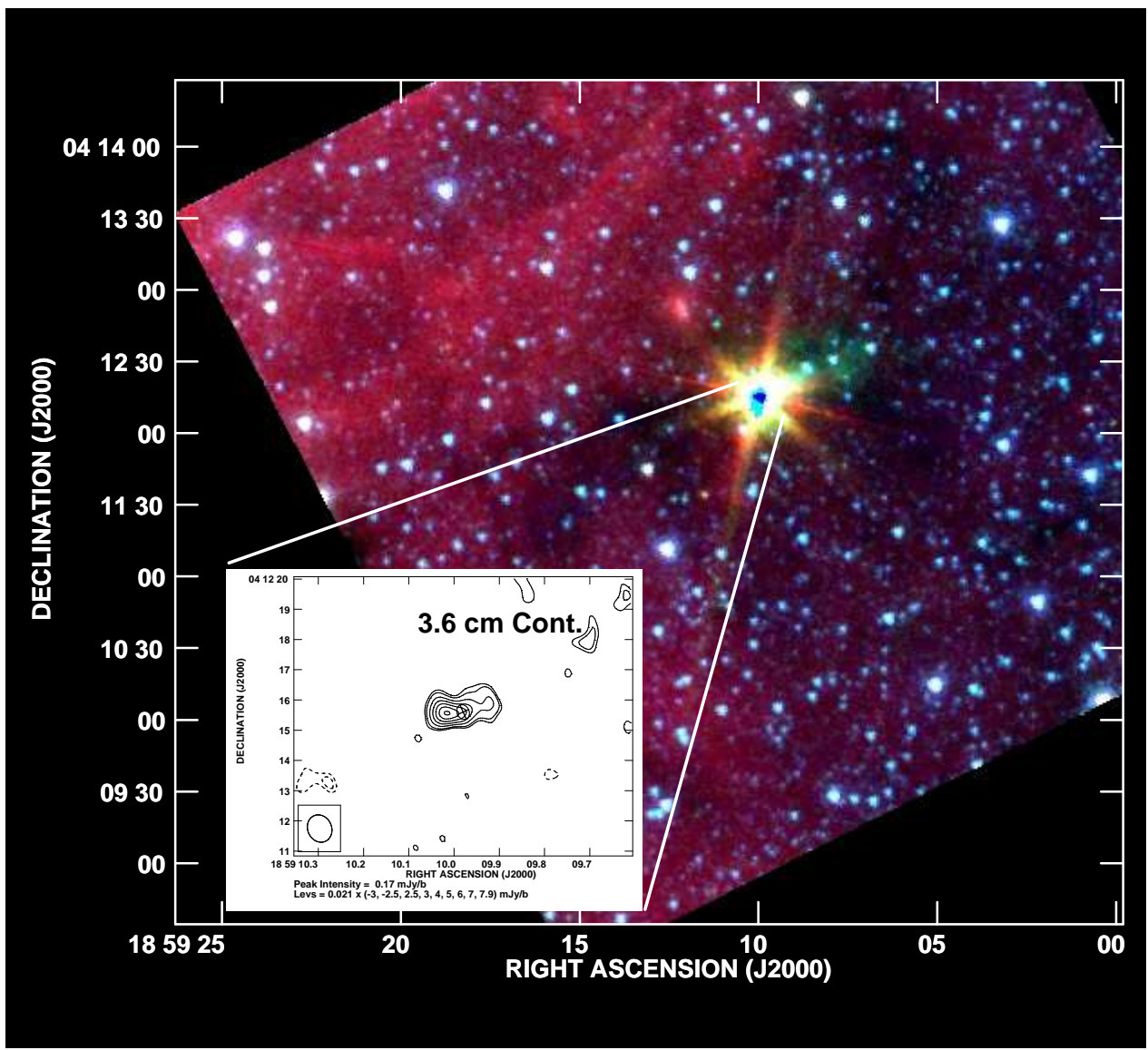}
\vspace*{-2cm}\caption{Three-color image of IRAS$\,$18566+0408 
from the Spitzer/IRAC 3.6$\,\mu$m (blue), 4.5$\,\mu$m (green), 
and 8.0$\,\mu$m (red) bands. The IRAC detectors were saturated at the 
IRAS$\,$18566+0408 peak.
Note the 4.5$\,\mu$m (green) excess toward the
NW of IRAS$\,$18566+0408. 4.5$\,\mu$m excess is a known
tracer of shocked gas in massive star 
forming regions (e.g., Smith et al. 2006).
In the inset we show the 3.6$\,$cm radio image obtained here
(see Figure~1). The elongation of the 3.6$\,$cm continuum 
source is approximately parallel to the 4.5$\,\mu$m NW excess, 
which agrees with the suggestion that the ionized gas traces
a jet in IRAS$\,$18566+0408.}
\label{f3}
\end{figure}

\clearpage

\begin{figure}
\hspace*{0cm}\vspace*{2cm}\includegraphics[scale=0.60, angle=-90]{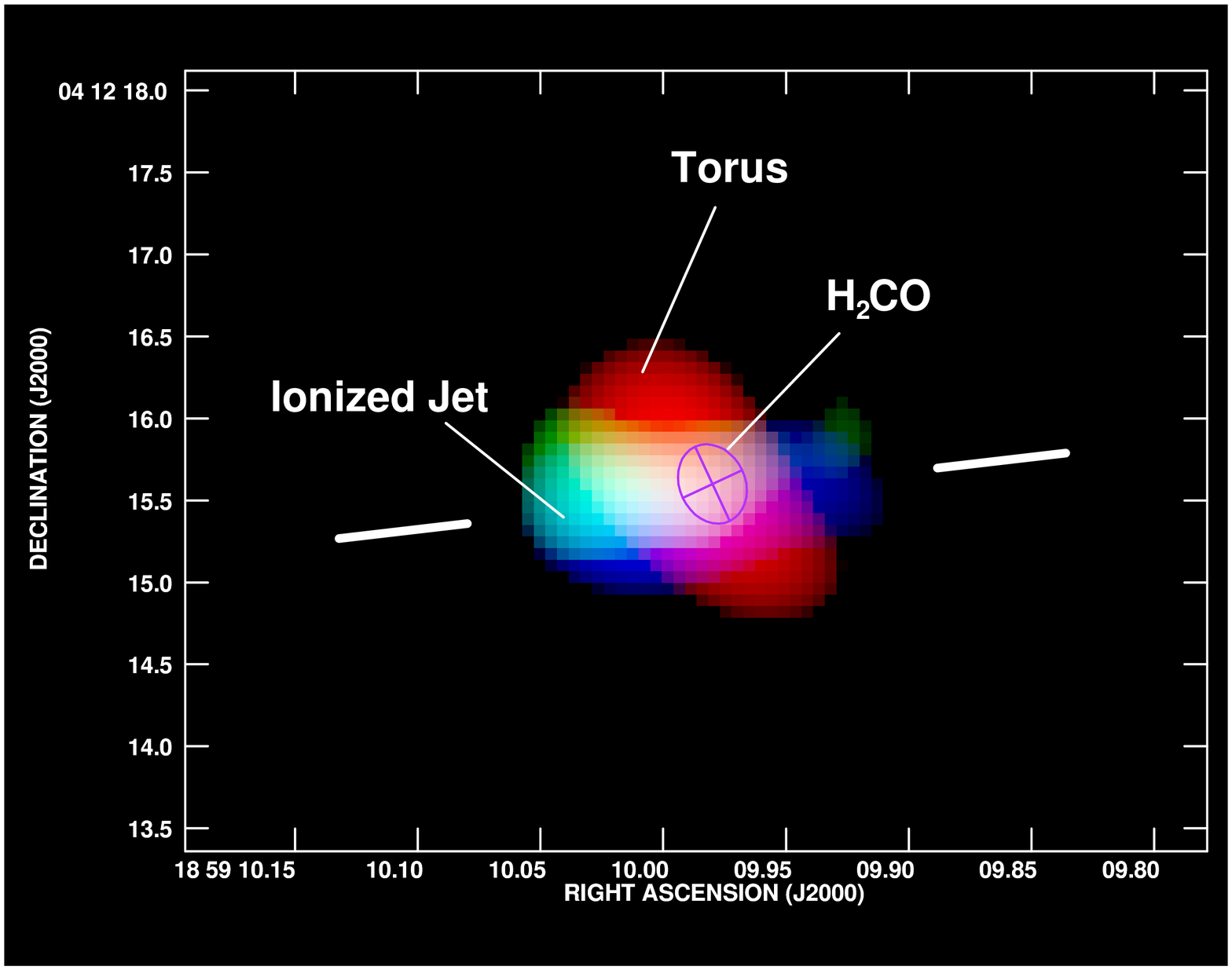}
\vspace*{-1.1cm}\caption{Three-color image of the 6$\,$cm (blue),
3.6$\,$cm (green), and 7$\,$mm (red) radio continuum emission from 
IRAS$\,$18566+0408. The colors were saturated to increase contrast. 
At cm wavelengths (where the emission is dominated by
ionized gas) the brightness distribution is oriented East-West (highlighted by
thick white lines), whereas at 7$\,$mm the orientation is
NE-SW. Given that the 7$\,$mm emission
is dominated by thermal dust and the ionized jet is almost perpendicular
to the 7$\,$mm source, then the 7$\,$mm flattened structure appears
to trace a circumstellar torus. 
The H$_2$CO maser position as detected by 
Araya et al. (2005) is shown by ``$\Earth$''. 
The maser is coincident with the center of the 7$\,$mm source.}
\label{f4}
\end{figure}


\begin{thebibliography}{}


\bibitem[Araya et al. (2006a)]{arayaG232006} Araya, E., Hofner, P.,
Goss, W. M., Kurtz, S., Linz, H., \& Olmi, L. 2006a, ApJ, 643, L33

\bibitem[]{745} Araya, E., Hofner, P., \& Goss, W. M. 2007a, 
in Astrophysical Masers and their Environments, ed. J. Chapman, 
\& W. A. Baan, (Cambridge: Cambridge Univ. Press), {\it in press}

\bibitem[Araya et al. (2007b)]{araya07_GBTVLA} Araya, E., Hofner, P., 
Goss, W. M., H. Linz, Kurtz, S., \& Olmi, L. 2007b, ApJS,
170, 152

\bibitem[Araya et al. (2005)]{araya05ir18566} Araya, E., Hofner, P.,
Kurtz, S., Linz, H., Olmi, L., Sewi{\l}o, M., Watson, C., \& Churchwell, E.
2005, ApJ, 618, 339

\bibitem[Araya et al. (2004)]{araya04arecibo} Araya, E., Hofner, P.,
Linz, H., Sewi{\l}o, M., Watson, C., Churchwell, E., Olmi, L., \&
Kurtz, S. 2004, ApJS, 154, 579

\bibitem[Araya et al. (2007c)]{flare} Araya, E., Hofner, P.,
Sewi{\l}o, M., Linz, H., Kurtz, S., Olmi, L., Watson, C., \& 
Churchwell, E. 2007c, ApJ, 654, L95

\bibitem[]{509} Arce, H. G., Shepherd, D., Gueth, F., 
Lee, C.-F., Bachiller, R., Rosen, A., \& Beuther, H. 
2007, Protostars \& Planets V, ed. B. Reipurth, D. Jewitt, \& K. Keil
(Tucson: Univ. of Arizona Press), 245

\bibitem[]{514} Beltr\'an, M. T., Cesaroni, R., Codella, C., 
Testi, L., Furuya, R. S., \& Olmi, L. 2006, Nature, 443, 28

\bibitem[]{517} Beuther, H., Churchwell, E. B., McKee, C. F., \&
Tan, J. C. 2007, Protostars \& Planets V,
ed. B. Reipurth, D. Jewitt, \& K. Keil
(Tucson: Univ. of Arizona Press), 165

\bibitem[Beuther et al. (2002a)]{beuther02a} Beuther, H., Schilke, P.,
Sridharan, T. K., Menten, K. M., Walmsley, C. M., \&
Wyrowski, F. 2002a, A\&A, 383, 892

\bibitem[Beuther et al. (2002b)]{beuther02b} Beuther, H., Walsh, A.,
Schilke, P., Sridharan, T. K., Menten, K. M., \&
Wyrowski, F. 2002b, A\&A, 390, 289

\bibitem[Boland \& de Jong]{boland81} Boland, W., \& de Jong, T. 1981,
A\&A, 98, 149

\bibitem[]{789} Bronfman, L., Nyman, L.-\AA., \& May, J. 1996,
A\&AS, 115, 81

\bibitem[Carral et al. (1999)]{carral99} Carral, P., Kurtz, S.,
Rodr\'{\i}guez, L. F., Mart\'{\i}, J., Lizano, S. \& Osorio, M.
1999, RevMexAA, 35, 97

\bibitem[]{537} Cesaroni, R., Galli, D., Lodato, G., 
Walmsley, C. M., \& Zhang, Q. 2007, in Protostars \& Planets V,
ed. B. Reipurth, D. Jewitt, \& K. Keil
(Tucson: Univ. of Arizona Press), 197

\bibitem[]{542} Churchwell, E. 2002, ARA\&A, 40, 27

%\bibitem[De Buizer \& Minier (2005)]{debuizer05} De Buizer, J. M.,
%\& Minier, V. 2005, ApJ, 628, L151

\bibitem[]{806} Fish, V. L. 2007, in Astrophysical Masers and Their 
Environments, ed. J. Chapman, \& W. A. Baan, (Cambridge: Cambridge
Univ. Press), {\it in press} (arXiv:0704.0242v1)

\bibitem[]{810} Forster, J. R., Goss, W. M., Gardner, F. F., 
\& Stewart, R. T. 1985, MNRAS, 216, 35

\bibitem[Garay et al. (2003)]{garay03} Garay, G.,  Brooks, K., Mardones, D., 
  \& Norris, R. P. 2003, ApJ, 587, 739

\bibitem[Garay et al. (2007)]{garay07} Garay, G., Mardones, D., Bronfman, L., 
Brooks, K. J.  Rodr\'\i guez, L. F., G\"usten, R., Nyman, L-{\AA}, 
Franco-Hern\'andez, R., \& Moran, J. M.  2007, A\&A, A\&A 463, 217

\bibitem[]{820} Hill, T., Thompson, M. A., Burton, M. G., Walsh, A. J., 
Minier, V., Cunningham, M. R., \& Pierce-Price, D. 
2006, MNRAS, 368, 1223

\bibitem[Hoffman et al. (2003)]{hoffman03} Hoffman, I. M., Goss, W. M.,
Palmer, P., \& Richards, A. M. S. 2003, ApJ, 598, 1061

\bibitem[Hoffman et al. (2007)]{hoffman07} Hoffman, I. M., Goss, W. M.,
\& Palmer, P. 2007, ApJ, 654, 971

\bibitem[]{557} Litvak, M. M. 1970, ApJ, 160, L133

\bibitem[]{559} Mart\'{\i}, J., Rodr\'{\i}guez, L. F., \& Reipurth,
B. 1995, ApJ, 449, 184

\bibitem[]{835} Mehringer, D. M., Goss, W. M., \& Palmer, P.
1995, ApJ, 452, 304

\bibitem[]{838} Miralles, M. P., Rodr\'{\i}guez, L. F., \& 
Scalise, E. 1994, ApJS, 92, 173

\bibitem[]{538} Molinari, S., Brand, J., Cesaroni, R., \& Palla, F.
1996, A\&A, 308, 573

\bibitem[]{844} Noriega-Crespo, A. et al. 2004, ApJS, 154, 352

\bibitem[]{565} Ossenkopf, V., \& Henning, Th. 1994, A\&A, 291, 943

%\bibitem[]{848} Pestalozzi, M. R., Elitzur, M., Conway, J. E.,
%\& Booth, R. S. 2004, ApJ, 603, L113 

\bibitem[Pratap et al. (1994)]{pratap94} Pratap, P., Menten, K. M.,
\& Snyder, L. E. 1994, ApJ, 430, L129

\bibitem[]{570} Reynolds, S. P. 1986, ApJ, 304, 713

\bibitem[]{572} Rodr\'{\i}guez, L. F., Zapata, L. A., \&
Ho, P. T. P. 2007, ApJ, 654, L143

\bibitem[]{575} Rodr\'{\i}guez, L. F., et al. 1998, Nature, 395, 355

\bibitem[]{577} Smith, H. A., Hora, J. L., Marengo, M., \&
Pipher, J. L. 2006, ApJ, 645, 1264

\bibitem[]{864} Sridharan, T. K., Beuther, H., Schilke, P., 
Menten, K. M., \& Wyrowski, F. 2002, ApJ, 566, 931

\bibitem[]{580} Thaddeus, P. 1972, ApJ, 173, 317

\bibitem[Zhang (2005)]{zhang05} Zhang, Q. 2005, in Massive Star Birth:
A Crossroads of Astrophysics, ed. R. Cesaroni, M. Felli, E. Churchwell,
\& M. Walmsley, (Cambridge: Cambridge Univ. Press), 135

\bibitem[]{873} Zhang, Q., Sridharan, T. K., Hunter, T. R., Chen, Y.,
Beuther, H., \& Wyrowski, F. 2007, A\&A, {\it in press} 
(arXiv: 0704.2767v1)


\end{thebibliography}
\end{document}